\documentstyle[aps,pre]{revtex}
\tighten   

\begin{document}

\title{Trapping Dynamics with Gated Traps: Stochastic Resonance-Like Phenomenon}
\author{Alejandro D. S\'anchez \thanks{
Present address: Instituto de F\'{\i}sica de
Cantabria, Fac. de Ciencias, Universidad de Cantabria, Santander,  Spain,
e-mail: asanchez@ifca.unican.es},
Jorge A. Revelli \thanks{Fellow CONICET, Argentina; e-mail: revelli@cab.cnea.gov.ar}
and Horacio S. Wio\thanks{ Member of CONICET, Argentina; e-mail:
wio@cab.cnea.gov.ar}}
\address{Grupo de F\'{\i}sica Estad\'{\i}stica 
\thanks{http://www.cab.cnea.gov.ar/CAB/invbasica/FisEstad/estadis.htm}\\
Centro At\'omico Bariloche (CNEA) and Instituto Balseiro (CNEA and UNC) \\
8400 San Carlos de Bariloche, Argentina}

\maketitle

\begin{abstract}
We present a simple one-dimensional trapping model prompted 
by the problem of ion current across biological membranes. 
The trap is modeled mimicking the ionic channel membrane 
behaviour. Such voltage-sensitive channels are  
open or closed depending on the value taken by a potential. Here 
 we have assumed that the external potential has two contributions:
 a determinist periodic and a stochastic one. 
Our model shows a resonant-like maximum when we plot 
the amplitude of the oscillations in the absorption current
vs. noise intensity. The model was solved both numerically
 and using an analytic approximation and was found to be in
 good accord with numerical simulations.
\end{abstract}

\vskip 1.truecm

\normalsize

During the last decade we have witnessed a growing interest in the
phenomenon of stochastic resonance (SR). This phenomenon is
related to the enhancement of the response of a nonlinear system
due to the interplay between nonlinear oscillations and noise. SR
has been studied in different physical, chemical and biological
contexts \cite{SR3}. In particular, it has been found to play a
relevant role in several problems in biology: mammalian sensory
systems, increment of the tactile capacity, visual perception,
low frequency effects and low amplitude electromagnetic fields,
etc \cite{biology}.

Among the many experiments showing SR there is one related to
the measurement of the current through voltage-sensitive ion
channels in a cell membrane \cite{nature}. These channels switch
randomly between open and closed states, controlling the ion
current. This and other related phenomena have motivated several 
theoretical studies of the problem of ionic transport through 
biomembranes. Those studies have used different approaches, as 
well as different forms of characterizing
stochastic resonance in such systems \cite{membrana}.

In this work we want to analyze the effect of the simultaneous action of a 
deterministic and a stochastic external field on the trapping rate of a 
gated imperfect trap.
Here we do not pretend to make a precise modeling
 of the behaviour of an ionic channel. We propose a simple model of dynamic trapping.
 This model is motivated by the above indicated experiment on cell membranes\cite{nature}.
Our main result is that even a simple model of a gated trapping process 
can show a SR-like behaviour.

The study of gated trapping processes, i.e. a trapping process where the 
traps have some kind of internal dynamic has attracted considerable 
interest \cite{varios}. Many authors discussed 
the way to link the gated trapping processes with the measured behavior 
of the so called ionic pumps \cite{membrana}. For example, among other
 factors, the ion transport depends on membrane electric potential
 (which plays the role of the barrier height) and can be stimulated by 
both {\it dc} and {\it ac} external fields. 

We based our study on the so called {\it stochastic model} for reactions
\cite{stoch,stoch1,stoch2}, that has been generalized in order to include 
the internal dynamics of traps. The dynamical process consists of the
opening or closing of the traps according to an external field.
 Such a field has two contributions, one periodic with a 
small amplitude, and the other stochastic whose intensity will be 
the tuning parameter.

Our starting model equation is
\begin{equation}
\label{model}
\partial_t \rho(x,t)=D \partial^2_x \rho(x,t)- \gamma(t) \delta(x)
\rho(x,t)+n_u,
\end{equation}
where $\gamma$ is a stochastic process that represents the absorption
probability of an individual trap in our system, 
$\rho$ is the particle density (particles that have not been yet trapped);  
for a given realization of $\gamma$, $x$ is the coordinate over the
one-dimensional system and $n_u$ is a source term that
represents a constant flux of ions. The injection of ions can be at
a trap position or at any other position. In this last case the
ion can diffuse to the trap position. This diffusion
coefficient would represent an effective diffusion through the volume
rather a diffusion over the membrane surface.

Each $\gamma$ is modelled as follows:
\begin{equation}
\label{gamma} \gamma(t)=\gamma^* \theta[B \sin (\omega
t)+\xi-\xi_c],
\end{equation}
where $\theta (x)$, the Heaviside function, determines when the trap is open 
or closed. The trap works in the following manner: if the signal composed of the 
harmonic part plus $\xi$ (the noise contribution) reaches a threshold $\xi_c$ 
the trap opens, otherwise it is closed.  We are interested in 
the case where $\xi_c>B$, that is, without noise the trap is always closed. 
When the trap is open the particles are trapped  with a 
given frequency (probability per unit time) $\gamma^*$. In other words the open trap 
is represented by  an ``imperfect trap" .  Finally, in order to complete the model, we
must give the statistical properties of the noise $\xi$. We assume
that $\xi$ is an uncorrelated Gaussian noise of width $\xi_0$, i.e.
\begin{equation}
\label{noise}
\langle \xi (t) \xi (u) \rangle= \delta_{t,u} \xi_0^2.
\end{equation}
Note that this is not a standard white noise \cite{vk} due to the
Kronecker symbol (instead of Dirac's delta $\delta(t-u)$).

For future reference we compute $\langle \gamma(t) \rangle$
(brackets mean averages over all realizations of the noise). The
result is
\begin{equation}
\label{gm}
\langle \gamma(t) \rangle=
\frac{\gamma^*}{2} {\rm erfc}\left[ \frac 1 {\sqrt{2} \xi_0}
(\xi_c-B \sin (\omega t))\right].
\end{equation}
Setting $n(x,t)=\langle \rho(x,t) \rangle$ and averaging Eq.
(\ref{model}) we can write
\begin{equation}
\label{unasola}
n(x,t)=n_u t-\int_0^t du {\it J}(u) G(x,t-u),
\end{equation}
where $G(x,t)=G_0(x,t)$ is the free diffusion propagator, and 
${\it J}(t)=\langle \gamma_j(t) \rho(j l,t) \rangle$ is the current
trough the trap. This current satisfies the relation
\begin{equation}
\label{current}
{\it J}(t)=\langle \gamma(t) \rangle \left [ n_u
t-\int_0^t du {\it J}(u) G(0,t-u)\right].
\end{equation}
We solve this equation numerically, using a standard ``trapezoidal" 
integral approach. 

The present problem can be extended to a
finite lattice with periodic boundary conditions.
In this case Eq. (\ref{model}) is replaced by
\begin{equation}
\label{modelm}
\partial_t \rho(x,t)=D \partial^2_x \rho(x,t)-\sum_{j=-\infty}
^{\infty} \gamma(t) \delta(x-j
l)+n_u,
\end{equation}
where, due to the boundary conditions, the sum runs over all the images of the trap.
The current satisfies the same Eq. (\ref{current}) but now
$G(x,t)=\sum_j G_0(x-j l,t)$.
This equation can be solved numerically.

In addition to the numerical solution, as in typical SR studies \cite{SR3},  
we can make some analytic calculations using a perturbation procedure. 
We assume $B/\xi_0$ to be small and expand
$\langle \gamma(t) \rangle$ (or $\langle \gamma_j(t) \rangle$) in
the following form
\begin{equation}
\label{serie}
\langle \gamma(t) \rangle=\langle \gamma_j(t) \rangle=\gamma_0+\gamma_1 \sin (\omega t)+{\cal O}((B/\xi_0)^2),
\end{equation}
where
\begin{eqnarray}
\gamma_0&=&\frac{\gamma^*}{2} {\rm erfc} \left( \frac{\xi_c}{\sqrt{2}\xi_0}\right)\\
\gamma_1&=&\frac{\gamma^* B}{\sqrt{2 \pi}\xi_0} \exp
\left(- \frac{\xi_c^2}{2 \xi_0^2}\right).
\end{eqnarray}
Now we can transform Eq. (\ref{current}) to the Laplace domain and
solve it iteratively. The expressions obtained are rather
complicated and for simplicity we present only the asymptotic
behavior. For an infinite lattice the asymptotic current is
\begin{equation}
{\it J}(t) \sim 4 n_u \sqrt{\frac{D t}{\pi}} +|C|\cos(\omega
t+\phi),
\end{equation}
where
\begin{equation}
C=|C| \exp(i \phi)=-\frac{i 4 \gamma_1 n_u \sqrt{ i \omega D t / \pi}}{\gamma_0
(\gamma_0/(2 \sqrt{D})+\sqrt{i \omega})},
\end{equation}
while for a finite one it is
\begin{equation}
{\it J}(t) \sim n_u l+|C|\cos(\omega t+\phi),
\end{equation}
where
\begin{equation}
C=|C| \exp(i \phi)=\frac{2 l \, n_u \omega \gamma_1 \sqrt{D}}{\gamma_0^2 \sqrt{i \omega}
\coth\left( \frac{l \sqrt{i \omega}}{2 \sqrt{D}} \right)+2 i \sqrt{D} \gamma_0 \omega}.
\end{equation}
The qualitative behaviour of the system changes since for
an infinite lattice the current grows without limit while in a finite
lattice the current reaches a stationary value when the source
particle supply is balanced with the trapping rate.

These expressions are valid
only for large values of the width noise $\xi_0$. In the opposite
case, i. e. for small values of $\xi_0$, the expansion of $\langle
\gamma \rangle$ (or $\langle \gamma_j \rangle$) as given in Eq.
(\ref{serie}) is not appropriate. In
this small $\xi_0$ limit we note that ${\it J}(t)$
is small and then we can obtain the following approximate
solution for Eq. (\ref{current})
\begin{equation}
{\it J}(t) \sim \langle \gamma (t) \rangle n_u t,
\end{equation}
that is independent of $G$ and hence is valid for both situations 
(finite and infinite lattice).
This expression is valid for short times such that we can
neglect the integral in Eq. (\ref{current}).

We choose to quantify the SR-like phenomenon by computing
the amplitude of the oscillating part of the absorption current
given by: 
\begin{equation}
{\Delta \it J}={\it J}|_{\sin(\omega t)=1} -{\it J}|_{\sin(\omega
t)=-1}
\end{equation}
 
The previous limits allow us to explain
the qualitative behaviour of the system as
follows. For a small noise intensity the current is low (remember
that $\xi_c>B$), hence ${\Delta \it J}$ is small too. For a large noise intensity the
deterministic part of the signal (the harmonic one) becomes
irrelevant and the ${\Delta \it J}$ is also small. Therefore, 
there must be a maximum at an intermediate value of the noise.  

We note that the case corresponding to a periodic array of fixed
traps only requires replacing $\gamma$ for $\gamma_j$ in the sum
of Eq. (\ref{modelm}). In the case of the $\gamma_j$ being independent of each
other, i.e. $\gamma_j(t)=\gamma^* \theta[B \sin (\omega
t)+\xi_j-\xi_c]$, where $\langle \xi_j (t) \xi_k (u) \rangle=
\delta_{j,k}\delta_{t,u} \xi_0^2$ the current within the SM is exactly
the same that for the finite lattice case.

The simulations were performed on one dimensional lattice of $L$ sites with
periodic boundary conditions. Initially there are no particles on
the lattice. The particles are injected randomly every $1/(n_u L)$ units of
time, with uniform distribution over the lattice, and are
allowed to perform a continuous time random walk (characterized by
the jump frequency at each neighbor site $q$). There is no restriction
on the number of particles at each site. A particle can be removed from the system
with a given probability distribution characterized by $\gamma$ when it
reaches the trap site. A detailed description of the 
algorithm used can be found in \cite{stoch2}. 
The reaction times were generated according to the following probability
density function
\begin{equation}
p(t)=\exp  \left( -\int _0 ^t \langle \gamma (t') \rangle dt' \right).
\end{equation}
All simulations shown in the figures correspond to averages over 1000 
realizations.
 
The results of both, numerical evolution and simulations, are shown
in Figs. 1 to 3.
 In Fig. 1 we show the temporal behaviour of the absorption current 
as a function of time. Figure 1{\it a} shows the evolution obtained
from numerical integrations  for a single trap (infinite system) and a 
periodic array of traps in a time interval such that the periodic array 
reaches its stationary behavior \cite{stoch1}. It is worth remarking that 
a system with 
a single trap does not have a stationary behaviour. However, the figure 
can induce the false impression that the single trap also presents a 
stationary behaviour. This is only an artifact of the short time period 
shown. The insert of Fig. 1{\it a} shows, in a much longer time scale, 
such a non stationary behaviour for the single trap problem. 
The agreement between both theoretical results makes it difficult to distinguish 
among them, particularly at short times. Figure 1{\it b} depicts the 
results obtained from the simulations (points) and the corresponding numerical 
integration of the single trap (solid line) for a fixed value of  $\xi_0$.
 One see the good agreement between the two curves. 
When looking at such curves, and for different
values of the noise intensity, it is apparent that for intensities smaller or 
larger than an {\it optimal} one, the amplitude is reduced, while in its  
neighborhood the amplitude becomes a maximum. This is better seen when 
plotting ${\Delta \it J}$, the amplitude of the oscillating part of the absorption current
  as a function of the noise intensity, as can be seen
in Fig. 2 for two different times. Such results make apparent the 
existence of a maximum as a function of $\xi_0$. 
Here we find good qualitative agreement between the results obtained by 
  numerical integration of Eq. (\ref{current}) and those from simulations. 
  The insert of Fig. 2 shows the results 
for the initial slope indicated in the figure by {\it h}, corresponding 
to the kind of curves shown in the principal part of the figure, as a function of time. 
That is, the slope of ${\Delta \it J}$ for low noise intensities and 
short times. Both results, obtained by numerical integration and simulations, show  
linear dependence with time as predicted in Eq. (15), however the actual values of the 
slopes are sightly different. 

 Finally in Fig. 3 we depict the result for $\phi$, the phase shift between the input periodic 
signal and the absorption current (as indicated in Eqs. (12) and (14)). We have 
compared the results from the numerical integration of Eq. (\ref{current}) and those 
from simulations, finding good agreement.

The present results show that even a simple trapping process with an adequate 
internal dynamic can present a SR like phenomenon. This 
suggests the possibility that such a kind of phenomenon could be ubiquitous in more
complex trapping like processes. 
                                                                                
A more elaborate model aimed at describing the experiment in Ref. \cite{nature},
including a more realistic representation of the ion channels 
is under study and will be the subject of further work.


{\bf Acknowledgments}

The authors want to thank to V. Gr\"unfeld for a critical reading
of the manuscript; and C. Budde and A. Bruschi for fruitful
discussions. Financial support from CONICET and ANPCyT (both
Argentine agencies) is acknowledged.

\begin{figure} 
\caption{Temporal evolution of the absorption  current of particles for a
gated trapping system. 
{\it a)} Time evolution obtained from numerical integration for a single
trap (solid line) and for a  finite system with periodic boundary 
conditions (points). We show the stationary behaviour for the periodic 
array of traps. 
The non stationary behaviour of the single trap problem, as discussed in 
the text, is apparent from the insert. 
{\it b)} Time evolution obtained from simulations (points), and from
numerical integration of a single trap system (solid line). 
The parameters are: $\omega=0.1$, $n_u=0.01$, $\xi_c=1.2$, $B=1.0$, 
$q_A=0.1$, $\gamma^{*}=10$, $\xi_{0}=0.3$ and $L=100$.}
\end{figure}

\begin{figure}
\caption{Variation of the final value of ${\Delta \it J}$ (amplitude of the oscillating part 
of the absorption current)  as a function of  $\xi_{0}$.
The results obtained from the numerical integration of Eq. (6) are indicated 
by solid lines, while those from simulations by circles.   
We  have chosen the following time parameters;  $a) t=517$ and $b) t=895$.
 The insert shows the initial slope {\it h} as a function of time.
The results obtained from the numerical integration are indicated by triangles
 and those from simulations by circles. The linear fitting is indicated by 
 continuous and dashed lines respectively. 
The results were obtained for low noise intensity values ($\xi_0\le0.1$).}  
\end{figure}


\begin{figure}
\caption{Phase shift $\phi$ as a function of $\xi_{0}$.
The results obtained for the numerical integration are indicated by a solid line
 while those from simulations by circles.}   
\end{figure}  

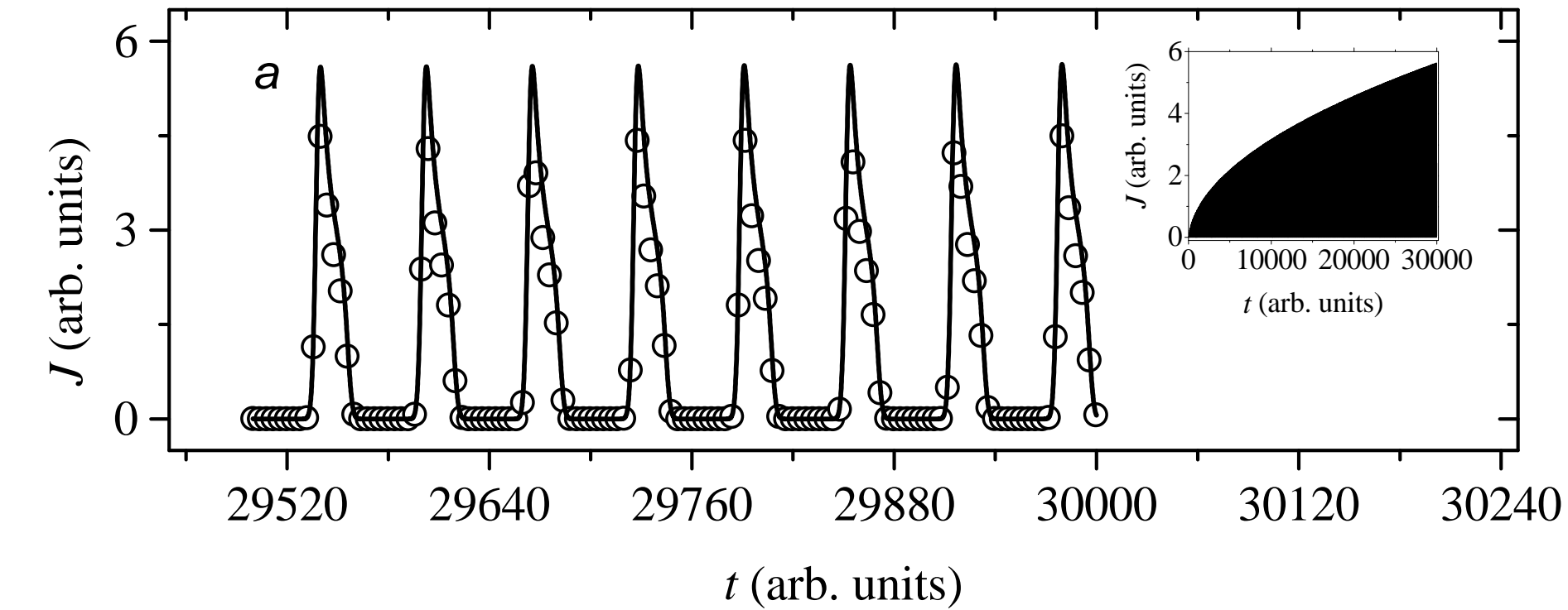
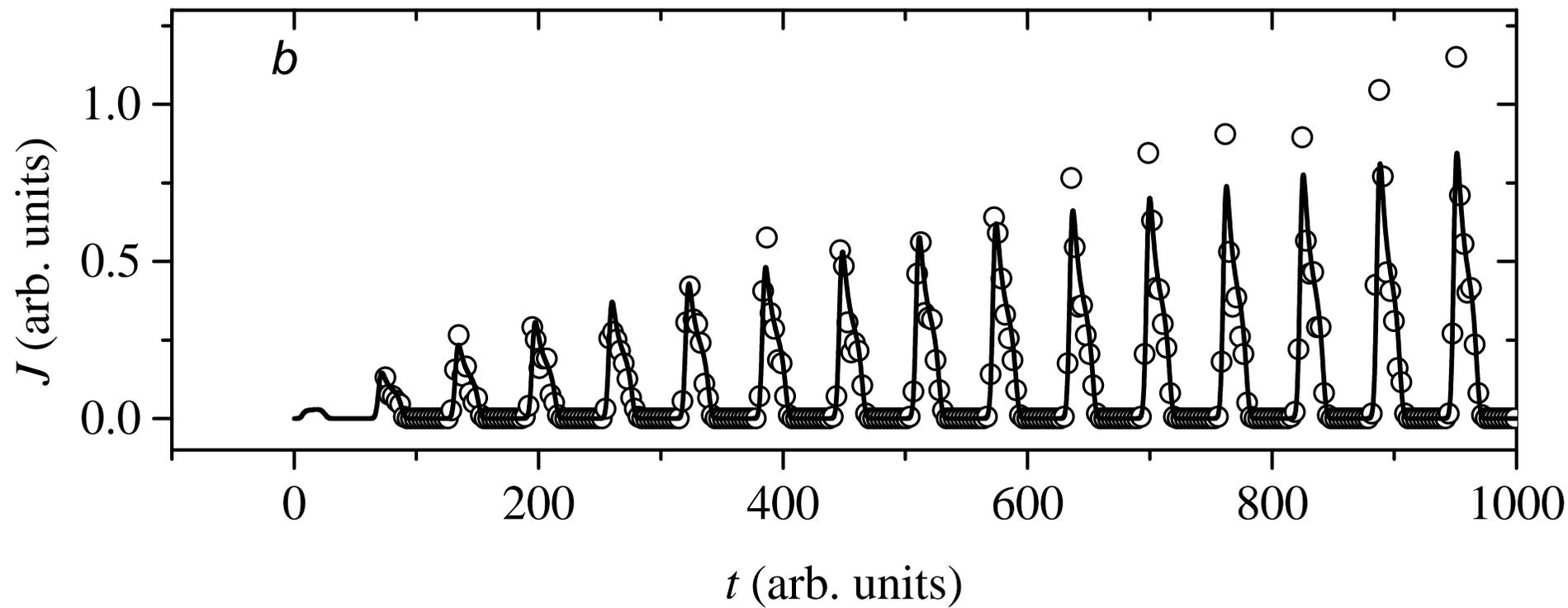
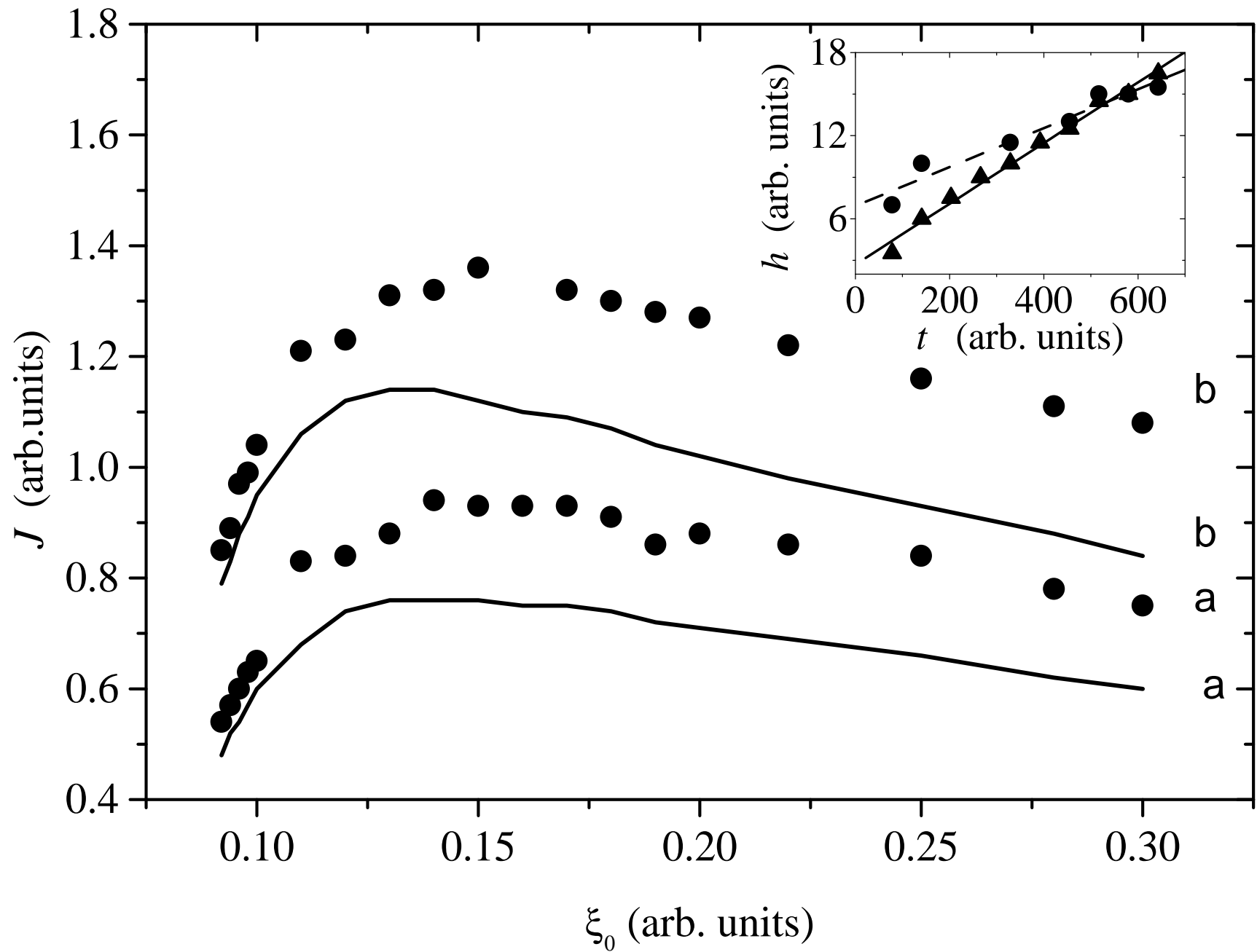
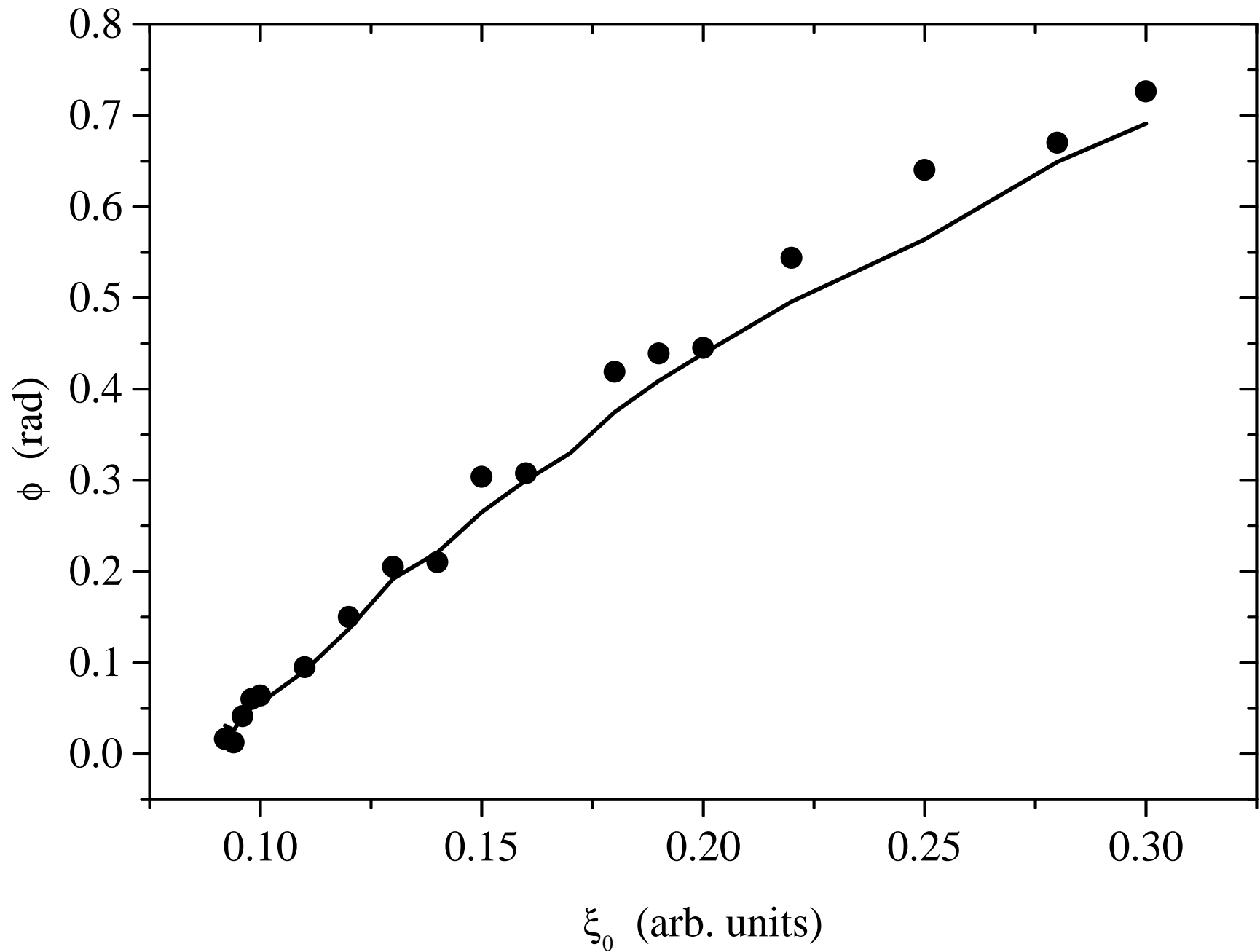
\end{document}